\documentclass[reprint,aps,pre,twocolumn,superscriptaddress,nofootinbib,floatfix]{revtex4-2}

\usepackage[T1]{fontenc}
\usepackage[utf8]{inputenc}
\usepackage{amsmath,amssymb,bm}
\usepackage{graphicx}
\usepackage{array}
\usepackage{microtype}
\usepackage{hyperref}

\hypersetup{colorlinks=true,linkcolor=blue,citecolor=blue,urlcolor=blue}

\newcommand{\Rbar}{\overline{R}}
\newcommand{\Calpha}{C_{\alpha}}
\newcommand{\etaw}{\eta_w}
\newcommand{\etatop}{\eta_{\rm top}}
\newcommand{\etafree}{\eta_{\rm free}}
\newcommand{\dB}{d_B}
\newcommand{\kappat}{\widetilde{\kappa}}

\begin{document}

\title{Cost convexity controls the architecture of resistance-optimal networks}

\author{Hugo Tovar}
\affiliation{Instituto Nacional de Medicina Genómica, Mexico City, Mexico}
\affiliation{Departamento de Física, Facultad de Ciencias, Universidad Nacional Autónoma de México, Mexico City, Mexico}
\author{Genaro Vázquez-Victorio}
\affiliation{Departamento de Física, Facultad de Ciencias, Universidad Nacional Autónoma de México, Mexico City, Mexico}
\author{Enrique Hernández-Lemus}
\affiliation{Instituto Nacional de Medicina Genómica, Mexico City, Mexico}
\author{Alejandra Cervera}
\email{acerverat@inmegen.gob.mx}
\affiliation{Instituto Nacional de Medicina Genómica, Mexico City, Mexico}
\author{Daniel Pérez-Calixto}
\email{daniel_perez@ciencias.unam.mx}
\affiliation{Instituto Nacional de Medicina Genómica, Mexico City, Mexico}
\affiliation{Departamento de Física, Facultad de Ciencias, Universidad Nacional Autónoma de México, Mexico City, Mexico}

\begin{abstract}
A conducting network improves access to its interior only by spending material,
so architectures compare only at equal budget. We redistribute conductance under
a power-law resource law $\Calpha=\sum_e a_e w_e^{\alpha}$ and measure access by
the reciprocal of the mean pairwise effective resistance. When the cost is
strictly convex the allocation is unique and every admissible edge keeps positive
conductance; at $\alpha=1$ uniqueness survives but an edge can be priced out. The
marginal worth of an edge is then its squared biharmonic distance, a Murray-type
balance whose balanced quantity is global rather than junction-local. The same
optimum scores any weighted network at its own budget, separating an allocation
factor from a topological one. Under concave cost the marginal price of an
incipient channel diverges, the feasible set is nonconvex, and conductance
condenses onto sparse backbones. The uniform star and the uniform complete graph
cross at $\alpha_c(n)=\log(n/2)/\log(n-1)$, near which intermediate states beat
both. Up to $n=64$ the best states found grow approximately linearly in $n$, so
channels per node change much less than the active fraction does. Loops are governed instead by a collapse
exponent $\alpha^{*}(n)$, resolved to $0.005$ at every size: it rises from
$0.555$ at $n=8$ to $0.675$ at $n=64$, and the collapse across it is abrupt, the
support at $n=64$ falling from $133$ active edges to the $63$ of a spanning tree.
A fixed exponent therefore changes character with size by crossing this boundary.
One prescribed source and sink instead gives a loopless optimum.
\end{abstract}

\maketitle

\section{Introduction}
Transport capacity and material cost are inseparable in a conducting network.
Raising an edge conductance improves access to the interior, and it consumes
resources that are then unavailable elsewhere. Any statement that one
architecture transports better than another is therefore incomplete unless the two are
held at the same budget.

Effective resistance is the natural measure of access when edge weights are
conductances. It accounts for parallel routes instead of collapsing transport
onto a single shortest path~\cite{Klein1993,Latora2001,Newman2010}, and its sum
over node pairs is the Kirchhoff index. Minimizing that sum under a linear budget
is a classical convex design problem with a well-understood
solution~\cite{Ghosh2008}. This leaves two questions. How does the optimal
architecture change when the physical price of conductance is nonlinear? And how
much of the available access does an arbitrary weighted network actually use?

Power-law prices arise whenever conductance is set by a channel geometry that
itself costs volume, surface, or mass, as in vascular, porous, thermal, and
electrical structures. The exponent is not a fitting convenience. Convex cost
penalizes concentration and rewards spreading conductance thinly; concave cost
does the reverse, and can drive the optimum onto a sparse support. Optimal
transport networks have been shown to pass between loopless and redundant
architectures as the cost law or the loading is
changed~\cite{Banavar2000,Bohn2007,Shao2007,Durand2006,Durand2007,Katifori2010,Ronellenfitsch2016}.
Those studies prescribe sources and sinks, or minimize dissipated power. The
loading here is different: every unordered node pair is weighted equally, so the
objective is the mean effective resistance and the network is optimized for
transport between any two sites rather than between selected terminals. The
distinction is inert in the convex regime and decisive in the concave one, where
the all-pairs objective can retain loops while the exact single-pair optimum
cannot.

The two cost regimes also require different standards of evidence. In the convex
case, the unique reference state makes the normalized efficiency well defined,
and the gradient shows where additional conductance provides the largest gain in
access. Below $\alpha=1$, several supports may compete. Multiple local minima
arise from the nonconvex formulation rather than from numerical noise, so the
reference can only be reported as the best state found within a stated search.
Over the sizes examined here, these states occupy a decreasing fraction of the
candidate graph even though their absolute support continues to grow. We
interpret this behavior through the constructal law, comparing each network with
the best configuration allowed by the same resource model~\cite{Bejan2004,Bejan2010}.

\section{Access at fixed resource}
\subsection{Conductance network and global access}
Consider a connected undirected graph $G=(V,E)$ with node set $V$, admissible edge
set $E$, $n=|V|$, and $m=|E|$. The non-negative edge conductances are collected in
$W=(w_e)_{e\in E}$. After fixing an arbitrary orientation for each edge,
$b_e\in\mathbb{R}^n$ is its incidence vector, carrying $+1$ and $-1$ at the
endpoints. The weighted Laplacian is
\begin{equation}
L(W)=\sum_{e\in E}w_e\,b_e b_e^{\mathsf T}.
\label{eq:laplacian}
\end{equation}
For nodes $s$ and $t$ with standard basis vectors $e_s$ and $e_t$, write
$b_{st}=e_s-e_t$. The effective resistance between them is
$r_{st}=b_{st}^{\mathsf T}L^{+}b_{st}$, with $L^{+}$ the Moore--Penrose
pseudoinverse of $L$. Averaging over the $n(n-1)/2$ unordered pairs gives
\begin{equation}
\Rbar(W)=\frac{2}{n-1}\operatorname{tr}(L^{+}).
\label{eq:Rbar}
\end{equation}
The prefactor $2/(n-1)$ converts the Kirchhoff index $n\operatorname{tr}(L^{+})$
into a pairwise mean, and it follows from $\sum_{i\neq j}r_{ij}=2n\operatorname{tr}(L^{+})$
together with the fact that the rows of $L^{+}$ sum to zero. Equation~(\ref{eq:Rbar})
represents a uniform ensemble of unit current injections and withdrawals. We take
global access to be $A(W)=1/\Rbar(W)$, the reciprocal of the mean resistance and
not the mean of the pair conductances.

\subsection{Resource law}
\label{sec:setup}
The price of conductance is
\begin{equation}
\Calpha(W)=\sum_e a_e w_e^{\alpha},\qquad a_e>0,\quad \alpha>0.
\label{eq:cost}
\end{equation}
The coefficient $a_e$ separates the candidate topology from the local price of an
edge, and $\alpha$ fixes how that price grows with conductance. In the spatial
examples below we use the length penalty $a_e=\ell_e$ with $\ell_e$ the Euclidean
channel length.

Both $\alpha$ and the length dependence of $a_e$ must be inherited from a physical
model rather than chosen for convenience. If $w_e$ is the end-to-end
Hagen--Poiseuille conductance of a tube, then $w_e\propto r_e^4/\ell_e$ and the
tube volume scales as $\ell_e^{3/2}w_e^{1/2}$, so a fixed-volume law corresponds to
$\alpha=1/2$ and $a_e\propto\ell_e^{3/2}$. Table~\ref{tab:mechanisms} collects the
pairs $(\alpha,a_e)$ that follow from the standard mechanisms in the same way. All
of them place the physically motivated exponent at or below the convex boundary,
which is why the concave regime is the relevant one rather than a mathematical
curiosity. The simpler choice $a_e=\ell_e$ used in most of the spatial examples
below is a controlled spatial penalty and is not offered as a universal
tube-volume law; the fixed-volume case itself is computed in
Sec.~\ref{sec:spatial}.

\begin{table}[b]
\caption{\label{tab:mechanisms}%
Cost exponents implied by standard channel mechanisms, obtained by expressing the
conserved resource in terms of the end-to-end conductance $w_e$ of a channel of
length $\ell_e$.}
\begin{ruledtabular}
\begin{tabular}{llll}
Channel & Conductance & Resource & $(\alpha,\,a_e)$\\
\hline
Viscous tube & $w\propto r^4/\ell$ & volume $\propto r^2\ell$ & $(1/2,\ \ell^{3/2})$\\
Viscous tube & $w\propto r^4/\ell$ & wall area $\propto r\ell$ & $(1/4,\ \ell^{5/4})$\\
Ohmic wire & $w\propto A/\ell$ & volume $\propto A\ell$ & $(1,\ \ell^{2})$\\
\end{tabular}
\end{ruledtabular}
\end{table}

At budget $C_0>0$ the reference allocation solves
\begin{equation}
\min_W \Rbar(W)\quad\text{subject to}\quad \Calpha(W)\le C_0,\qquad W\ge0,
\label{eq:optimization}
\end{equation}
with $W\ge0$ understood componentwise. A weighting whose positive-conductance
support is disconnected is assigned $\Rbar=+\infty$. Because $\Rbar$ strictly
decreases when any edge conductance is raised, every finite optimum exhausts the
budget, $\Calpha(W)=C_0$.

A minimizer exists for every $\alpha>0$, including the concave range where the
feasible set is not convex. The constraint $\Calpha(W)\le C_0$ with $a_e>0$ bounds
each coordinate by $w_e\le(C_0/a_e)^{1/\alpha}$, so the feasible set is compact;
$\Rbar$ is continuous on the subset of weightings whose support is connected and
diverges as connectivity is lost, so the infimum is attained in the interior of
that subset whenever the candidate graph is connected. Uniqueness is a separate
question and is guaranteed only for $\alpha\ge1$.

\subsection{Scoring a network against its own optimum}
The optimum of Eq.~(\ref{eq:optimization}) is useful beyond its own value, because
it converts a resistance into an efficiency:
\begin{equation}
\eta_\alpha(W\mid\Omega)=\frac{\Rbar^{*}_{\Omega}\!\left(\Calpha(W)\right)}{\Rbar(W)}.
\label{eq:eta}
\end{equation}
Here $\Omega$ is the admissible edge set and $\Rbar^{*}_{\Omega}(C)$ is the
minimum of $\Rbar$ over $\Omega$ at budget $C$. The set may contain only the edges
of an observed network, or all unordered node pairs. For $\alpha\ge1$ and a fixed
candidate edge set, Eq.~(\ref{eq:optimization}) is convex and $\Rbar$ is strictly
convex on connected weightings, so the optimal allocation is unique and
$0<\eta_\alpha(W\mid\Omega)\le1$, with equality only at that allocation. We
refer to this ratio as the constructal efficiency index.

For $0<\alpha<1$, the global reference generally cannot be obtained from an
uncertified numerical minimum. We therefore retain $\eta_\alpha$ only where the
global optimum is known. Elsewhere we report the resistance of the best state
found, its support, and its dependence on initialization; using a local minimum
as the denominator would give an apparently absolute score without the required
guarantee.

\section{The marginal worth of an edge}
\label{sec:marginal}
The effect of a single edge can be obtained without solving the optimization
problem at all. Let $\mathbf{1}$ be the all-ones vector. As long as the conducting
support stays connected, the nullspace of $L$ is $\operatorname{span}\{\mathbf{1}\}$
and $\delta L^{+}=-L^{+}(\delta L)L^{+}$ on the orthogonal subspace
$\mathbf{1}^{\perp}$. Perturbing $w_e$ then gives
\begin{equation}
\frac{\partial\Rbar}{\partial w_e}
=-\frac{2}{n-1}\,b_e^{\mathsf T}(L^{+})^2 b_e
=-\frac{2}{n-1}\,\dB^2(e).
\label{eq:gradient}
\end{equation}
Equation~(\ref{eq:gradient}) itself is not new: the derivative
$\partial\operatorname{tr}L^{+}/\partial w_e=-\|L^{+}b_e\|_2^2$ is standard in the
convex resistance-design literature~\cite{Ghosh2008}, and its identification with a
biharmonic distance follows Refs.~\cite{Lipman2010,Black2024}. What is new here is
Eq.~(\ref{eq:kappa}) below, its division by the $\alpha$-dependent marginal price,
and the boundary criterion at $w_e=0$ that this division produces.

The squared biharmonic distance is $\dB^2(e)=b_e^{\mathsf T}(L^{+})^2b_e=\|L^{+}b_e\|_2^2$~\cite{Lipman2010,Black2024},
the squared norm of the node-potential response to a unit current driven across
$e$. It is not the square of the effective resistance of that edge. The two
correlate strongly and differ at order one: on
\texttt{gnp\_random\_graph(14,\,0.35,\,seed=4)}, which has 36 edges, with
$w_e\sim U(0.3,1.0)$, they correlate at $0.990$, yet substituting $r_e^2$ for
$\dB^2(e)$ misstates the gradient by up to $89\%$ and reorders the ranking of
interventions. The substitution is literal, $-[2/(n-1)]\,r_e^2$ in place of
$-[2/(n-1)]\,\dB^2(e)$: the two have different units, and a proxy rescaled to a
common mean would give a smaller discrepancy.

Because $A=1/\Rbar$, the marginal access gain carries an additional factor
$\Rbar^{-2}$. Dividing by the marginal price of the same edge defines the gain per
unit cost,
\begin{equation}
\kappat_e\equiv\frac{\partial A/\partial w_e}{\partial\Calpha/\partial w_e}
=\frac{2\,\dB^2(e)}{(n-1)\,\Rbar^{2}\,\alpha\,a_e\,w_e^{\alpha-1}}.
\label{eq:kappa}
\end{equation}
At a stationary point that exhausts the budget, the Karush--Kuhn--Tucker
conditions equalize $\kappat_e$ across all active edges. This network-level
balance is a Murray-type statement, in that it equates a marginal transport
benefit with a marginal material cost~\cite{Murray1926,Lonardi2021}, but the
analogy becomes physical only once the dependence of $w_e$ on channel geometry
and dissipation is specified.

Durand reached a related result along a different route~\cite{Durand2006}:
minimizing global flow resistance at fixed channel volume or surface constrains
the cross sections of channels meeting at a junction, again generalizing Murray's
law. Here the constraint acts on conductances rather than directly on channel
geometry, so $\alpha$ belongs to the assumed material law and is not fixed by the
flow regime. The marginal quantity in Eq.~(\ref{eq:kappa}) is also different: it
is a biharmonic distance that depends on the whole network, not only on one
junction. A local balance is recovered only when the two descriptions coincide.

The boundary $w_e=0$ is what separates the two regimes, and it does so through
Eq.~(\ref{eq:kappa}) rather than through any property of $\Rbar$. For $\alpha>1$
the marginal price $\alpha a_e w_e^{\alpha-1}$ vanishes at $w_e=0$, so an
admissible edge with finite $\dB$ cannot stay absent at the optimum: its marginal
opening cost vanishes while its transport benefit remains positive. For $0<\alpha<1$ the
same marginal price diverges as $w_e\to0$, which suppresses the opening of a new
channel and makes sparse supports locally stable. At $\alpha=1$ the marginal price is the constant
$a_e$, neither vanishing nor divergent, and the split between zero and positive
conductances is decided by the complementarity conditions of the convex problem.
An edge stays closed whenever $\dB^2(e)/a_e$ falls below the common value that the
active edges share, so the linear case admits sparse optima even though it is
convex and its solution unique.\footnote{Three nodes suffice. Take the triangle
with $a_{01}=a_{02}=1$ and $a_{12}=a$, unit budget and $\alpha=1$. For $a<3$ all
three edges carry conductance. At $a=3$ the edge $(1,2)$ closes, and for $a\ge3$
the optimum is the path $1\!-\!0\!-\!2$ with $w_{01}=w_{02}=1/2$ and
$\Rbar=8/3$. The threshold is exact: at that state $\dB^2(1,2)=8$ and
$\dB^2(0,1)=\dB^2(0,2)=8/3$, so the closure condition
$\dB^2(1,2)/a\le\dB^2(0,1)/1$ first binds at $a=3$. The linear case is the classical
problem of Ref.~\cite{Ghosh2008}, where zero weights are a generic feature of the
solution.}

\section{The convex optimum as an absolute reference}
When every unordered node pair is admissible and $a_e=1$, the candidate graph is
complete. Permutation symmetry together with strict convexity gives the uniform
conductance
\begin{equation}
w_0=\left[\frac{2C_0}{n(n-1)}\right]^{1/\alpha}
\label{eq:w0}
\end{equation}
as the unique optimum for $\alpha\ge1$, and substitution into
Eq.~(\ref{eq:Rbar}) gives
\begin{equation}
\Rbar^{*}_{\rm free}=\frac{2}{n}\left[\frac{n(n-1)}{2C_0}\right]^{1/\alpha}.
\label{eq:Rfree}
\end{equation}
A fixed candidate topology supplies a second reference. Writing
$\Rbar^{*}_{\rm fixed}$ for its minimum resistance at the same budget, the
efficiency relative to the free reference separates into two factors,
\begin{equation}
\etafree=\frac{\Rbar^{*}_{\rm free}}{\Rbar^{*}_{\rm fixed}}
         \cdot\frac{\Rbar^{*}_{\rm fixed}}{\Rbar}
        =\etatop\,\etaw,
\label{eq:factor}
\end{equation}
which defines $\etatop=\Rbar^{*}_{\rm free}/\Rbar^{*}_{\rm fixed}$ and
$\etaw=\Rbar^{*}_{\rm fixed}/\Rbar$. The topological factor measures how much
access the wiring pattern forfeits; the allocation factor measures how well the
available conductance is distributed on that pattern. The factorization is exact
because both numerator and denominator are evaluated at the same budget, and it is
what makes the two questions separable at all.

Table~\ref{tab:convex} evaluates both factors for a set of standard graphs at
common $n$, $\alpha$, and budget, and the two columns move independently. The
star, the ring, and the complete graph are edge-transitive, so their automorphism
group acts transitively on the variables being optimized and convexity forces the
optimum to be uniform, giving $\etaw=1$. Edge-transitivity, not
vertex-transitivity, is the property this argument needs: the star has a hub and
is not vertex-transitive. Their $\etatop$ nonetheless spans an order of
magnitude. The barbell shows the converse and explains why an edge count is not
a transport measure: it carries more edges than either random graph in the table,
but most of them lie inside its two dense cliques, and all intergroup transport
passes through a three-edge bridge. Its allocation factor is the lowest in the
table, $\etaw=0.373$, because uniform weights waste budget on clique edges that
cannot relieve the bridge.

\begin{table}[b]
\caption{\label{tab:convex}%
Convex reference factors for selected networks at $n=24$ and $\alpha=2$. $\etaw$
scores the allocation of conductance given the topology, $\etatop$ scores the
topology against the complete graph, and $\etafree=\etatop\etaw$ is the overall
efficiency at the network's own budget. The budget is each network's own,
$C_0=\sum_e w_e^{\alpha}$ at unit weights, and $\Rbar$ is compared at that budget
using $\Rbar(cW)=\Rbar(W)/c$. The random graphs are
\texttt{gnm\_random\_graph(24,\,72,\,seed=1)},
\texttt{barabasi\_albert\_graph(24,\,4,\,seed=1)} and
\texttt{watts\_strogatz\_graph(24,\,4,\,0.5,\,seed=1)}; the remaining entries are
\texttt{grid\_2d\_graph(6,\,4)}, \texttt{star\_graph(23)},
\texttt{barbell\_graph(11,\,2)}, \texttt{cycle\_graph(24)} and
\texttt{path\_graph(24)}. Since $\alpha\ge1$, no support truncation is applied.}
\begin{ruledtabular}
\begin{tabular}{lrrrr}
Network & Edges & $\etaw$ & $\etatop$ & $\etafree$\\
\hline
Complete & 276 & 1.000 & 1.000 & 1.000\\
Erd{\H{o}}s--R\'enyi & 72 & 0.934 & 0.398 & 0.372\\
Barabási--Albert ($m=4$) & 80 & 0.878 & 0.397 & 0.348\\
Watts--Strogatz ($p=0.5$) & 48 & 0.979 & 0.294 & 0.288\\
Square lattice ($6\times4$) & 38 & 0.986 & 0.195 & 0.193\\
Star & 23 & 1.000 & 0.151 & 0.151\\
Barbell & 113 & 0.373 & 0.194 & 0.072\\
Ring & 24 & 1.000 & 0.068 & 0.068\\
Path & 23 & 0.968 & 0.036 & 0.035\\
\end{tabular}
\end{ruledtabular}
\end{table}

\section{Where concentration begins to pay}
\begin{figure}[t]
\centering
\includegraphics[width=\columnwidth]{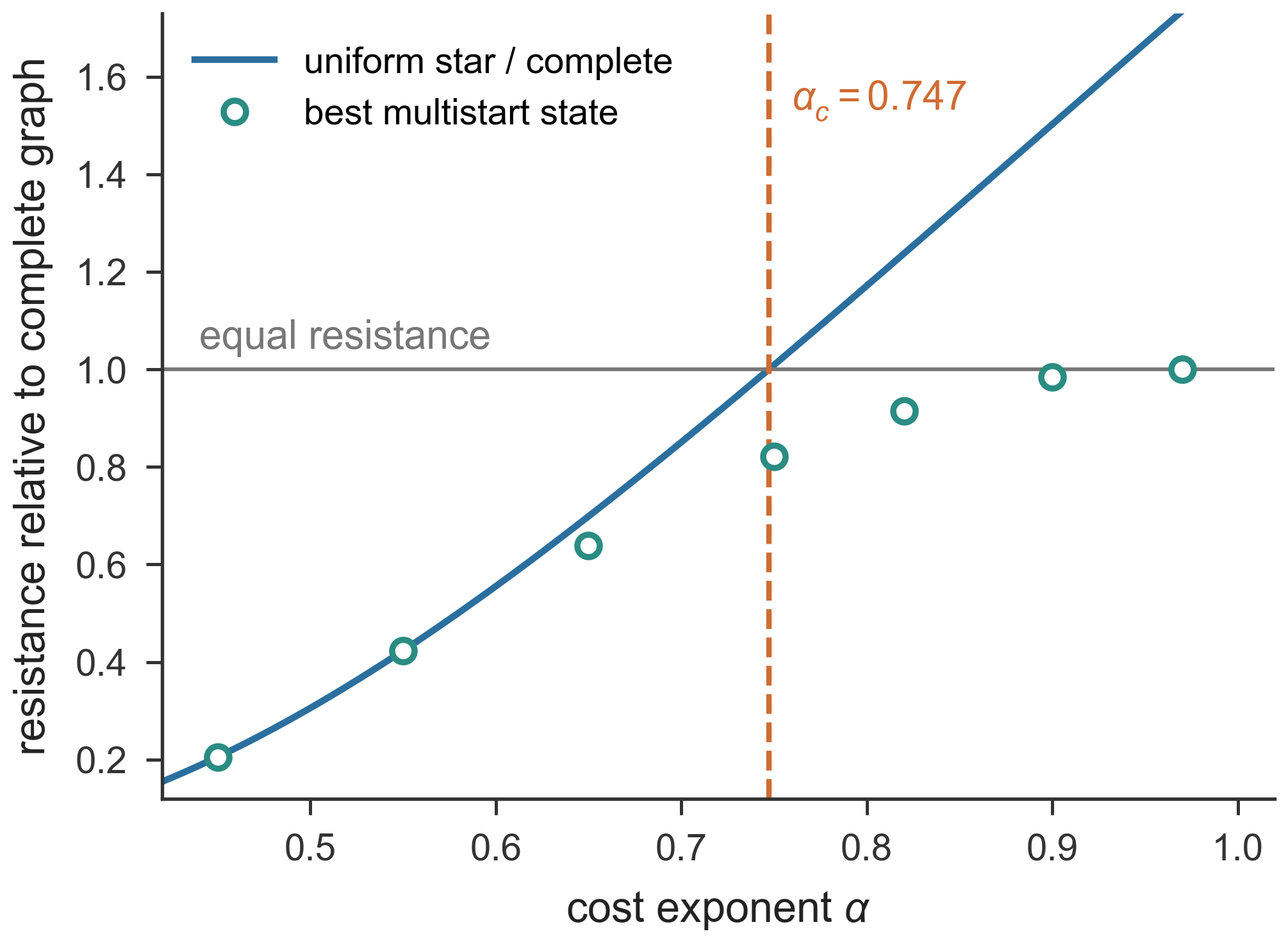}
\caption{\label{fig:starcomplete}%
Uniform star against uniform complete graph at $n=12$ and $C_0=1$. Open circles
give the lowest-resistance state obtained by multistart optimization, normalized
by the complete-graph resistance at the same $\alpha$. At the sampled values $\alpha\le0.55$ the state found coincides with the star. Near the analytic
crossing, intermediate supports beat both symmetric endpoints.}
\end{figure}

For $0<\alpha<1$ the complete graph remains a symmetric stationary state but need
not be optimal, and the simplest way to see where it loses is to race it against
the opposite extreme. At unit budget the uniform star and the uniform complete
graph have resistance ratio
\begin{equation}
\frac{\Rbar_{\rm star}}{\Rbar_{\rm complete}}=(n-1)\left(\frac{2}{n}\right)^{1/\alpha},
\label{eq:starcomplete}
\end{equation}
which equals one at
\begin{equation}
\alpha_c(n)=\frac{\log(n/2)}{\log(n-1)},
\label{eq:alphac}
\end{equation}
giving $\alpha_c=0.747$ at $n=12$. This is a crossing between two symmetric
ansätze and nothing more. Figure~\ref{fig:starcomplete} shows what the crossing
misses: near $\alpha_c$ the best state found by multistart search has lower
resistance than either endpoint, so it is neither symmetric endpoint.

\begin{figure*}[t]
\centering
\includegraphics[width=\textwidth]{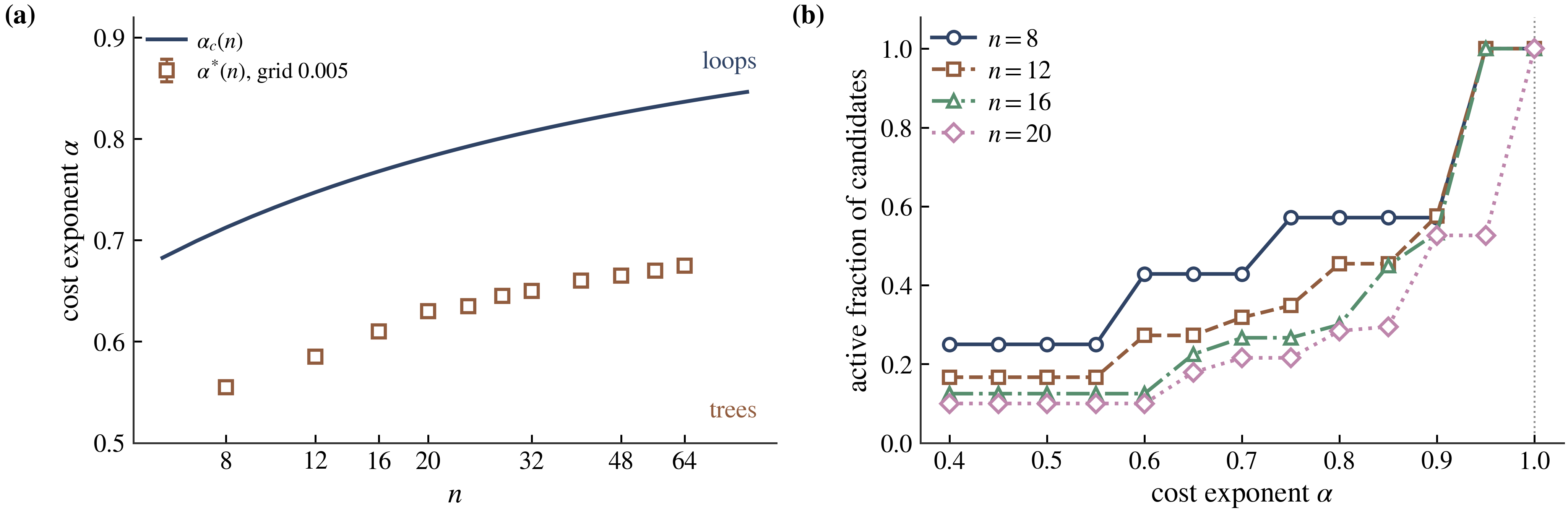}
\caption{\label{fig:finitesize}%
Finite-size behavior of the sparse regime. (a) The analytic crossing
$\alpha_c(n)=\log(n/2)/\log(n-1)$ (solid curve) against the loop-collapse
exponent $\alpha^{*}(n)$, the largest exponent whose best state is a spanning
tree, resolved at every tested size on a grid of step $0.005$ by the downward
sweep of Sec.~\ref{sec:numerics}. Error bars are one grid step. Both curves rise
with $n$, so the tree regime widens with system size; $\alpha^{*}$ lies well
below $\alpha_c$, which compares only the two symmetric ansatze and therefore
does not locate the boundary. (b) Fraction of candidate edges carrying
conductance in the state found for the four sizes in the finer scan. Below
$\alpha=1$ the support occupies a smaller share of the candidate set as $n$
grows.}
\end{figure*}

What the crossing does capture is a trend in system size. Since $\alpha_c(n)$
increases toward one, the interval in which the star beats the complete graph
widens as the network grows: $\alpha_c=0.712$, $0.747$, $0.768$, and $0.782$ at
$n=8$, $12$, $16$, and $20$, and $0.851$ at $n=100$. Figure~\ref{fig:finitesize} shows that the
measured boundary moves in the same direction. Resolving the largest tree-like
exponent on a grid of step $0.005$ gives $\alpha^{*}=0.555$, $0.585$, $0.610$, $0.630$, $0.635$, $0.645$, $0.650$,
$0.660$, $0.665$, $0.670$ and $0.675$ at $n=8$, $12$, $16$, $20$, $24$, $28$, $32$,
$40$, $48$, $56$ and $64$. The boundary rises at every size, as $\alpha_c(n)$ does, but lies
between $0.15$ and $0.16$ below it throughout: the two symmetric
ansatze bracket the mechanism without locating it. Sparse backbones are therefore
not an artifact of small systems. How far this trend goes is the subject of
Sec.~\ref{sec:scaling}.

\section{Numerical procedure}
\label{sec:numerics}
Below $\alpha=1$ it is convenient to work with the resource share of an edge,
\begin{equation}
p_e=\frac{a_e w_e^{\alpha}}{C_0},\qquad p_e\ge0,\qquad \sum_e p_e=1,
\label{eq:shares}
\end{equation}
from which the conductance is recovered as
\begin{equation}
w_e=\left[\frac{C_0\,p_e}{a_e}\right]^{1/\alpha}.
\label{eq:wefromp}
\end{equation}
The budget constraint is then automatic and the search runs on the simplex.

The allocation search has two stages. The shares are first optimized in softmax
coordinates with L-BFGS-B. Edges with $p_e\le\epsilon_{\rm cut}$, with
$\epsilon_{\rm cut}=10^{-5}$ throughout, are then set exactly to zero and the
shares on each remaining connected support are refined with BFGS, which removes
any residual dependence on nearly vanishing conductances. The truncation stage is
applied for $\alpha\le1$: only for $\alpha>1$ does Sec.~\ref{sec:marginal} give
every admissible edge positive conductance at the optimum, so truncating
there could only discard legitimately small weights. Feasibility of a support is
decided topologically rather than from the algebraic connectivity, so that a
connected support carrying uniformly small conductances is not misread as
disconnected.

Each exponent is initialized from a uniform state, a star, a minimum spanning
tree, a depth-first tree, and 60 seeded Dirichlet allocations, for 64 starts in
total; the finite-size scan of Fig.~\ref{fig:finitesize} and the size scan of
Sec.~\ref{sec:scaling} use the same four structured states with 24 random starts,
and the convex spatial cases use 16.

Equal numbers of starts at different sizes is not equal search quality. From
$n=8$ to $n=64$ the number of allocation variables grows by a factor of $72$,
while the fraction of starts reaching the best value is already between one and
five in $64$ at $n=12$. We therefore add a continuation stage: at fixed $n$ the
exponent is swept downward and the winning state at one exponent is supplied as an
additional start at the next, and the better of the blind and continued results is
kept. The outcome can only improve on blind multistart, since the blind starts are
retained. Measured this way the blind protocol is the weaker of the two, and the states it
misses differ structurally and not only in resistance. At $n=32$ it returns
$\beta=42$ at $\alpha=0.75$ and $\beta=27$ at $\alpha=0.66$, where continuation
reaches $\beta=53$ and $\beta=33$ at resistances lower by $8.5\times10^{-4}$ and
$5.8\times10^{-3}$ relative. The discrepancy in cycle rank is not one-signed: at
$\alpha=0.68$ the blind state carries one cycle \emph{more} than the better
state, $\beta=34$ against $33$, at a higher resistance. What the comparison
establishes is therefore not a bias of fixed sign but that a cycle rank read off
a blind search is a property of the state that search happened to find, and can differ from the best available state by as much as eleven independent
loops. On the fine grid of Sec.~\ref{sec:scaling} the same weakness appears as a
systematic offset: at seven of the eleven sizes the blind protocol returns a
spanning tree one grid step above $\alpha^{*}$, where continuation still holds a
looped state, so a collapse exponent read from blind search alone would sit one
step too high.

Every state reported below, including the size scan of
Table~\ref{tab:scaling} and Fig.~\ref{fig:scaling}, uses the continued protocol.
Continuation improved on blind multistart at five of the 44 grid points: four at
$\alpha=0.75$, where it raises the cycle rank from $5$ to $9$ at $n=8$, from $64$
to $83$ at $n=48$, from $83$ to $93$ at $n=56$ and from $88$ to $114$ at $n=64$,
lowering the resistance by $1.6$, $5.6$, $10.8$ and $9.8\times10^{-3}$ relative;
and one at $\alpha=0.65$ and $n=24$, from $19$ to $25$ at a gain of
$6.6\times10^{-3}$. It never improved on the blind result at $\alpha=0.55$, where
every state is a spanning tree, or at $\alpha=0.90$.

We checked the implementation in both coordinate systems. Central differences in
the conductances reproduce Eq.~(\ref{eq:gradient}) with a maximum relative
discrepancy of $2.5\times10^{-8}$. The optimizer works in logit coordinates, so
we checked that composite gradient separately after projecting out the
$\mathbf{1}$ direction left invariant by the softmax. Its maximum relative error
is $9.0\times10^{-8}$. For every state reported below, the budget residual
$|\Calpha(W)-C_0|$ vanishes to machine precision. The largest infinity norm of
the feasible projected gradient in the bounded logit coordinates is
$3\times10^{-6}$.

The dispersion across starts is a practical measure of nonconvexity, and it is
substantial. Counting a start as reaching the best state when its resistance lies
within $10^{-6}$ relative to the smallest value found, between one and five of 64
starts qualify at each exponent on the complete candidate graph. With $R_{\min}$
and $R_{\max}$ the smallest and largest final resistances, the relative spread
$(R_{\max}-R_{\min})/R_{\min}$ is $3.87$ at $\alpha=0.45$, stays above one through
$\alpha=0.82$, and falls below one at $\alpha=0.90$ and $0.97$; the interquartile
range divided by $R_{\min}$ falls from $0.092$ to $0.003$ across the same
interval. Every support transition reported below therefore refers to the best
state obtained at a stated search budget, not to a certified global minimum. The
controls described below focus on the conclusions most sensitive to this
distinction.

The calculations begin with the complete graph on 12 nodes and $a_e=1$. Spatial
effects are examined on a random geometric graph with 28 nodes,
connection radius $0.38$, seed 9, 95 edges, and $a_e=\ell_e$. The loading
comparison of Sec.~\ref{sec:loading} uses a second geometric graph with 20 nodes,
connection radius $0.42$, seed 5, and 62 edges. These seeded graphs are controlled
test beds and not empirical data sets.

\section{From backbone to mesh at fixed topology}
To isolate the effect of cost convexity we hold the candidate topology, the
budget, and the all-pairs loading fixed, and vary only $\alpha$. The candidate
graph is $K_{12}$, with 66 admissible edges and $a_e=1$.

Table~\ref{tab:beststates} reports the lowest-resistance allocation found at each
exponent. The strongly concave states are sharply localized: at $\alpha=0.45$ and
$0.55$ the best support has 11 active edges and agrees with the uniform star to
numerical precision. It grows to 18 edges at $\alpha=0.65$, 36 at $\alpha=0.90$,
and all 66 at $\alpha=0.97$. The effective support tracks the active-edge count
closely, which says the shares on the support are not far from uniform, while only
one to five starts in 64 reach the best value at any exponent.

\begin{table}[b]
\caption{\label{tab:beststates}%
Best multistart states on the complete candidate graph at $n=12$, from 64 starts.
Effective support is $1/\sum_e p_e^2$ and an edge counts as active when
$p_e>10^{-5}$. The last two columns give the fraction of starts reaching the best
value and the relative range of final resistances across starts.}
\begin{ruledtabular}
\begin{tabular}{cccccc}
$\alpha$ & $\Rbar_{\rm best}/\Rbar_{\rm complete}$ & Active & Eff. & Starts & Spread\\
&& edges & support & at best & \\
\hline
0.45 & 0.205 & 11 & 11.0 & 3.1\% & 3.87\\
0.55 & 0.423 & 11 & 11.0 & 3.1\% & 1.36\\
0.65 & 0.638 & 18 & 18.0 & 1.6\% & 1.48\\
0.75 & 0.821 & 23 & 22.8 & 1.6\% & 1.79\\
0.82 & 0.914 & 26 & 25.6 & 1.6\% & 2.08\\
0.90 & 0.984 & 36 & 35.4 & 1.6\% & 0.53\\
0.97 & 1.000 & 66 & 66.0 & 7.8\% & 0.73\\
\end{tabular}
\end{ruledtabular}
\end{table}

\begin{figure}[t]
\centering
\includegraphics[width=\columnwidth]{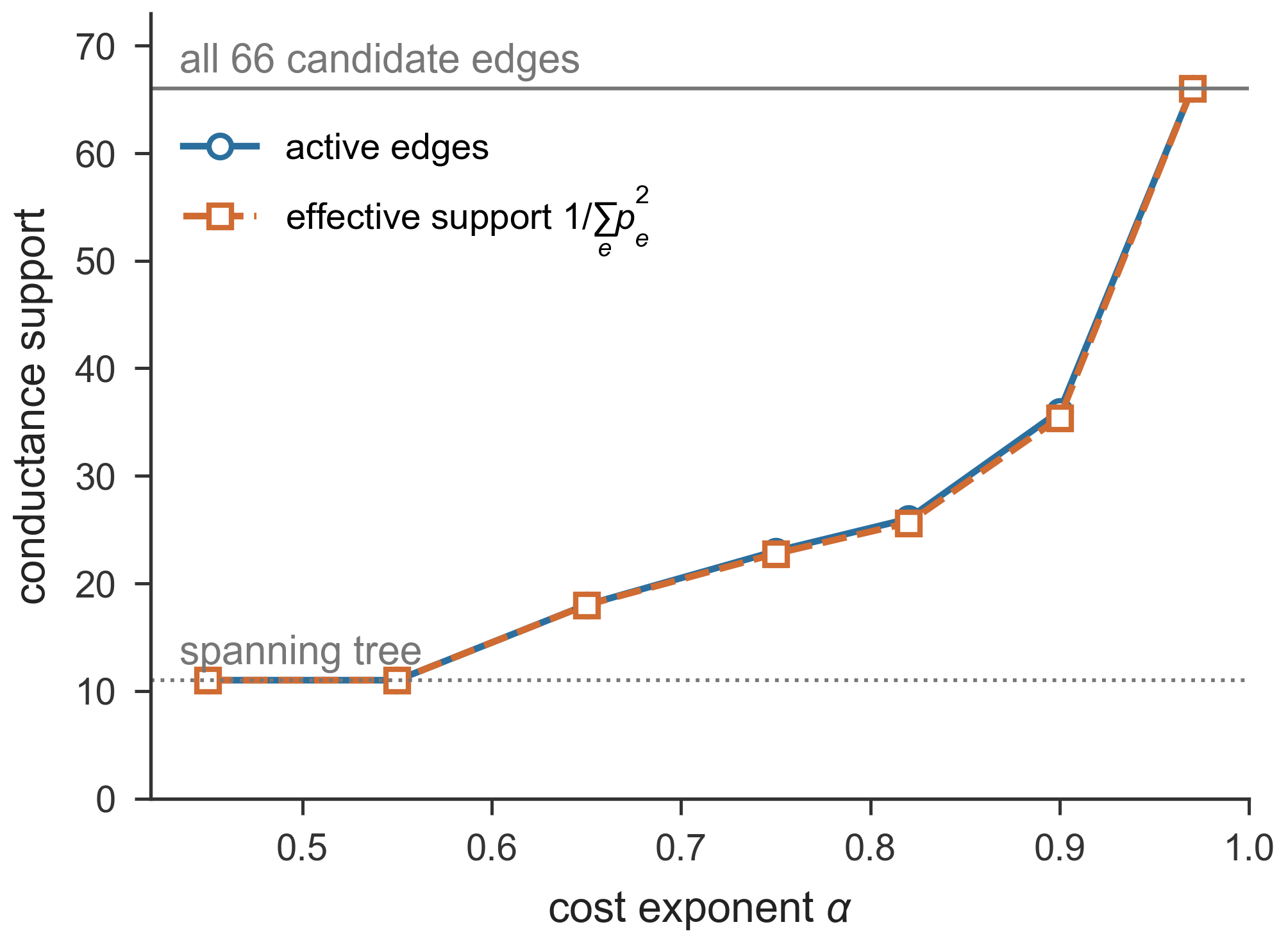}
\caption{\label{fig:support}%
Conductance support of the best multistart state on the complete candidate graph.
The tree value $n-1$ is marked by a dotted line. The support stays tree-like at
small $\alpha$ and expands in steps as the cost becomes less concave.}
\end{figure}

Figure~\ref{fig:support} shows the same progression as a structural sequence. The
active-edge count and the effective support sit at the tree value for
$\alpha=0.45$ and $0.55$, then rise in steps as $\alpha$ approaches one, and near $\alpha=0.97$ the search returns the complete graph. This approach to full
support is consistent with the convex limit at $\alpha=1$, but the steps themselves should not be
read as a thermodynamic critical sequence: the system is finite, the continuation
in $\alpha$ is discrete, and the objective is nonconvex with genuinely competing
minima. What is robust is the direction of travel, from a sparse backbone to a
distributed support.

\section{Finite-size growth of the active support}
\label{sec:scaling}
The preceding results concern relatively small graphs. On a complete candidate
graph the number of available edges grows as $n^2$, so the active fraction alone
can give a misleading picture of sparsity. We instead examine the absolute
support size. A dense state contains $O(n^2)$ active edges, whereas a backbone
contains $O(n)$. Distinguishing these trends requires larger systems.

We fix four concave exponents and increase $n$ to 64. At the largest size, the
complete candidate graph has 2016 allocation variables. The search budget remains
at 28 starts throughout the grid, keeping the comparison between sizes on equal
footing. For a connected active support the cycle rank is
$\beta=m_{\rm active}-n+1$, the number of independent loops.
Table~\ref{tab:scaling} summarizes the behavior at each exponent and reports the
largest-size values. Figure~\ref{fig:scaling} shows the complete size dependence;
all 44 numerical states are listed in Supplemental Table S1.

\begin{table*}[t]
\caption{\label{tab:scaling}%
Summary of the size scan on the complete candidate graph, $a_e=1$. Values in
the last four columns correspond to the largest size, $n=64$. All entries are
best multistart states, not certified global optima. The complete grid of 44
$(\alpha,n)$ combinations is reported in Supplemental Table S1. Active edges
are counted at $p_e>10^{-5}$ and $\beta$ is the cycle rank.}
\begin{ruledtabular}
\begin{tabular}{clrrrr}
$\alpha$ & behavior over the tested sizes & Active & $\beta$ &
$\mathrm{Active}/(n-1)$ & $\beta/n$ \\
\hline
0.55 & tree at all 11 sizes          &  63 &   0 & 1.00 & 0.00 \\
0.65 & tree for $n\geq32$            &  63 &   0 & 1.00 & 0.00 \\
0.75 & rising per-node ratio         & 177 & 114 & 2.81 & 1.78 \\
0.90 & looped; settles after $n=40$  & 300 & 237 & 4.76 & 3.70 \\
\end{tabular}
\end{ruledtabular}
\end{table*}

\begin{figure*}[t]
\centering
\includegraphics[width=\textwidth]{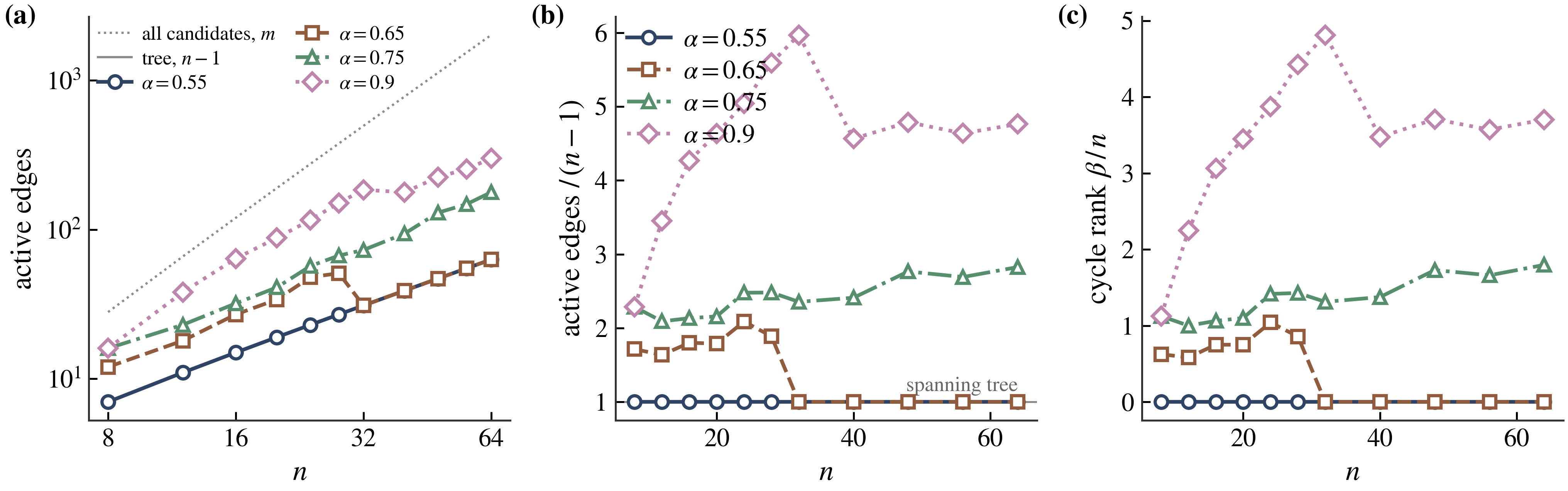}
\caption{\label{fig:scaling}%
Growth of the support in the concave states found. (a) Active edges against $n$ on
logarithmic axes, bracketed by the spanning-tree line $n-1$ and the full candidate
set $m=n(n-1)/2$. At the larger sizes the trends are approximately linear in $n$,
and every series stays far below the candidate line. (b) Active edges per node,
which varies little at the larger sizes, with a level set by $\alpha$; the
$\alpha=0.90$ series is still drifting below $n=40$. (c) Cycle rank per node,
which behaves in the same way, and drops to zero at $\alpha=0.65$ between $n=28$
and $n=32$.}
\end{figure*}

Over the range examined the states found grow approximately
linearly in $n$.
At $\alpha=0.75$ the ratio $\mathrm{active}/(n-1)$ fluctuates about $2.67$ over the
last four sizes; at $\alpha=0.90$ the same ratio is near $4.69$ over the last four.
We quote these directly measured per-node ratios rather than a fitted slope. A
straight-line fit over $n\ge20$ returns a slope that need not agree with the local
per-node ratio, because the fit also carries an intercept, and with only four points
at the largest sizes the two estimators are not distinguishable in any case. The active
fraction of the candidate set decreases roughly as $1/n$, which is not an
independent observation but the algebraic consequence of linear growth with
$m=n(n-1)/2$. At $\alpha=0.75$, the product
$n\,(\mathrm{active}/m)$ stays between $4.7$ and $5.6$ from $n=24$ to $64$, while
$m$ grows by a factor of seven. At $\alpha=0.55$ no fit is needed: the state found is a spanning tree, with $n-1$ edges and $\beta=0$, at all eleven sizes.
The normalized counts $\mathrm{active}/(n-1)$ and $\beta/n$ support a limited but
direct conclusion. Over the accessible range, the number of channels per node
changes much less than the fraction of the candidate graph that is used. Whether
that per-node count remains bounded as $n\rightarrow\infty$ is not settled here,
and the complete candidate graph does not test how it depends on the density of
admissible edges.

These finite sizes do not establish an asymptotic growth law. Over $8\le n\le64$ a
linear trend cannot be separated from $n^{1.08}$ or from $n\log n$: the effective
power-law exponents at $\alpha=0.75$ and $0.90$ are $1.215$ and $0.955$, and with
eleven sizes spanning less than a decade no model comparison among these
candidates is meaningful. We therefore rely on the directly
measured per-node ratios rather than assigning an asymptotic exponent to this
finite range.

Every active-edge count and cycle rank in this work inherits the cut
$p_e>10^{-5}$, and it is worth being precise about what we have and have not
tested there. Recounting the reported states at $10^{-4}$, $10^{-5}$ and
$10^{-6}$ leaves the active-edge count and the cycle rank identical at all three
cuts for $(\alpha,n)=(0.55,32)$, $(0.65,28)$, $(0.65,32)$, $(0.75,32)$,
$(0.75,64)$, $(0.90,32)$ and $(0.90,64)$. Taken alone that invariance would be
expected rather than informative, since the support-refinement step of
Sec.~\ref{sec:numerics} sets the discarded shares to exactly zero and a recount
afterwards cannot disagree with the cut that produced the support. We therefore ran the control the objection calls for. The search is repeated once
with truncation disabled, and the refinement is rerun from that same pretruncation
solution at each cut, so the active-edge count and the cycle rank are genuine
functions of $\epsilon_{\rm cut}$. They do not move: at all three cuts the seven
states give $31$, $51$, $31$, $73$, $151$, $185$ and $300$ active edges with
$\beta=0$, $24$, $0$, $42$, $88$, $154$ and $237$, and the refined resistances
agree to every printed digit. What also speaks to the
question is the margin, since across these states the smallest retained share
ranges from $2.0\times10^{-3}$ to $3.2\times10^{-2}$, two to four orders of
magnitude above the loosest cut, so nothing sits near $10^{-5}$ for a different
choice to reclassify. The threshold-free effective support agrees closely as
well, $31.0$ against 31 active edges at $(0.55,32)$ and $294.3$ against 300 at
$(0.90,64)$.

Loops behave differently from the support that carries them, and the $\alpha=0.65$
row of Table~\ref{tab:scaling} is where this shows. The best support found carries loops
up to $n=28$, with $\beta/n$ between $0.58$ and $1.04$ and no clean trend, and is an exact
spanning tree at every one of the five larger sizes, $n=32$, $40$, $48$, $56$, and
$64$. Within this search budget the best states found therefore change character with
system size at fixed cost exponent: looped states win at $n\le28$ and none was
found that beats the tree at any larger size. This is an apparent numerical
transition in the states we can locate, and it concerns more than
Eq.~(\ref{eq:alphac}), which compares only the two symmetric ansätze. Whether the
true optimum changes character at some size between $28$ and $32$ is not
established here.

Because the apparent loss of loops could result from incomplete sampling, we
performed two additional checks. First, we raised the blind search budget from 28
to 64 and then 144 starts. The best resistance at $\alpha=0.65$ remains unchanged
in every printed digit, at
$381.631590$ for $n=32$ and $546.797241$ for $n=40$, a relative change of exactly
zero. Since blind Dirichlet sampling explores 500 to 800 dimensions poorly, the
second check follows the looped state directly. We take the 51-edge, 24-cycle state found at $n=28$,
carry it into $K_{32}$ and $K_{40}$ by retaining its shares and attaching each new
node to the busiest node of the old support, and use that as the initial condition.
The seeded state relaxes onto the same spanning tree, with a resistance identical
to the blind result. The looped state that wins at $n=28$ is therefore not
simply overlooked at $n=32$ and $40$.

Both checks vary the sampling. Neither reaches the move that actually opens a
loop, and for $\alpha<1$ that move is not infinitesimal: the marginal price
$\alpha a_e w_e^{\alpha-1}$ diverges as $w_e\rightarrow0$, so no gradient step
ever creates a channel from nothing, and raising the number of blind starts cannot
supply what the search does not contain. Two further controls address this
directly.

The third carries a looped state \emph{downward in} $\alpha$ at fixed $n$, the
direction complementary to the seeded continuation above, on a grid of step
$0.005$. At $n=32$ the continued state holds at $64$ active edges with $\beta=33$
at every exponent from $\alpha=0.680$ down to $\alpha=0.655$, and at
$\alpha=0.650$ it falls to $31$ edges with $\beta=0$ and $\Rbar=381.631590$, the
value the blind search returns. The collapse is thus abrupt rather than gradual,
and it is abrupt at every size: the transition is single at all eleven, the
largest tree-like and smallest looped exponent differing by exactly one grid
step, and the cycle rank immediately above it is $2$, $7$, $10$, $18$, $18$, $30$,
$33$, $41$, $54$, $62$ and $70$ from $n=8$ to $n=64$. At $n=64$ the support
falls from $133$ active edges to the $63$ of a spanning tree across a step of
$0.005$ in $\alpha$. A discontinuous onset of loops has been reported for optimal
supply networks under prescribed loading, where it arises through a saddle-node
bifurcation as the cost of new edges is lowered~\cite{Kaiser2020}; the collapse
seen here is the same phenomenon viewed along the cost exponent.

Resolving $\alpha^{*}(n)$ also reframes the $\alpha=0.65$ row of
Table~\ref{tab:scaling}. That row is not a place where loops stop being worth
buying at large $n$; it is a horizontal line in the $(\alpha,n)$ plane crossing a
boundary that rises with $n$. Since $\alpha^{*}=0.635$ and $0.645$ at $n=24$ and $28$ but $0.650$ at $n=32$, the fixed exponent $0.65$ sits above the boundary at
the two smaller sizes and at or below it from $n=32$ on, which is exactly the
change of character the row reports. Read that way the apparent size-driven loss
of loops is a crossing rather than a mechanism, and it predicts where the same
crossing occurs at other exponents: $\alpha=0.60$ would cross between $n=12$ and
$16$, and $\alpha=0.68$ has not crossed by $n=64$.

The fourth tests the finite move itself. Taking the tree found at
$(\alpha,n)=(0.65,32)$, each of the $465$ closed edges in turn is assigned a finite
share $\varepsilon$, the remaining shares are renormalized, and the state is
refined. At $\varepsilon=0.01$, $0.02$ and $0.05$ no insertion improves on the
incumbent: all $465\times3$ trials relax back to the same tree, the best recovered
resistance differing from it by less than $2\times10^{-12}$ in relative terms. The
tree is thus stable against opening any single channel at these scales. Varying the
random start set at fixed $(\alpha,n)=(0.65,32)$ over three independent seeds also
returns the same state in every printed digit.

These are robustness checks and not a proof: they were run at these sizes only,
they probe single-edge insertions rather than simultaneous ones, and they cannot
exclude some other looped basin, at these or at larger $n$, that beats the tree.
Certification of global optimality is not available for this problem.

One entry in Table~\ref{tab:scaling} does move with the search budget, and we
report it rather than smooth it. At $\alpha=0.90$ and $n=40$, going from 28 to 144
starts lowers the best resistance by $1.4\times10^{-4}$ in relative terms and
shifts the support from 178 to 179 active edges. The $\alpha=0.90$ series is the
densest tested and sits closest to the convex boundary, where the number of
competing supports is largest, and its per-node ratio rises monotonically to
$5.97$ at $n=32$ before falling to the plateau near $4.7$. We read the sizes below
$n=40$ there as pre-asymptotic rather than as evidence against saturation, and we
report them so the reader can judge that call.

\section{A length-weighted spatial network}
\label{sec:spatial}
The complete candidate graph has no geometry, so the same calculation is repeated
with the length penalty $a_e=\ell_e$ on a fixed random geometric graph. Alongside
the resource effective support $N_p=1/\sum_e p_e^2$ we now also record the
conductance effective support $N_w=1/\sum_e q_e^2$ with $q_e=w_e/\sum_j w_j$. The
two differ because Eq.~(\ref{eq:wefromp}) is nonlinear, so a concentrated budget
can still produce a broad conductance profile, and the gap between them is itself
informative.

\begin{figure*}[t]
\centering
\includegraphics[width=\textwidth]{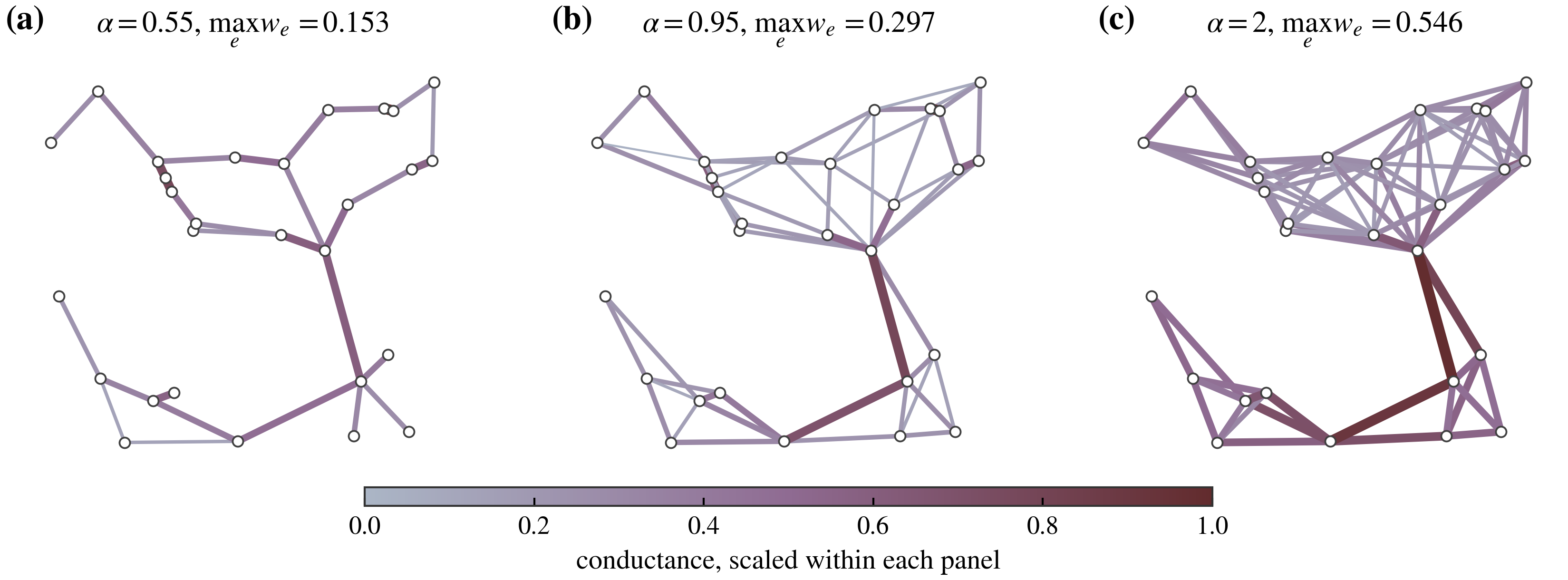}
\caption{\label{fig:spatialstates}%
Best multistart conductance states in the spatial candidate network at
(a) $\alpha=0.55$, (b) $\alpha=0.95$, and (c) $\alpha=2$. Line width and color
encode conductance, scaled within each panel, with the panel maximum
$\max_e w_e$ given in each title so that the three are comparable; edges with
$p_e\le10^{-5}$ are omitted. The support changes from a sparse backbone to a loop-rich mesh as the
cost law moves from the concave to the convex regime.}
\end{figure*}

The progression of Table~\ref{tab:spatial} and Fig.~\ref{fig:spatialstates}
matches the dimensionless case. At $\alpha=0.55$ the best state has 30 active
edges and $\beta=3$: a near-tree backbone carrying three independent cycles, on a
28-node graph whose spanning trees have 27 edges. Five cycles are present at
$\alpha=0.75$; by $\alpha=0.95$, 60 edges are active and $\beta=33$; and the two
strictly convex cases activate every candidate edge, as Sec.~\ref{sec:marginal}
requires. This example uses one realization of the geometric graph and is a
controlled test bed rather than a sample from an ensemble, so it shows that
geometry is compatible with the same progression, not how the transition depends
on geometry.

The linear boundary is the sharpest test of Sec.~\ref{sec:marginal}. At
$\alpha=1$ the problem is still convex and the optimum unique, but the marginal
price $a_e$ no longer vanishes at $w_e=0$, so an edge whose biharmonic distance
does not pay for its length can be priced out. With $a_e=\ell_e$ this is exactly
what happens: $86$ of the $95$ candidate edges are active and nine are excluded,
between the 60 of $\alpha=0.95$ and the full support of $\alpha=1.25$. Sparsity at
the convex boundary is therefore not a numerical artifact of the concave regime; it
is the complementarity predicted by Eq.~(\ref{eq:kappa}), realized on a graph with
unequal prices rather than on the three-node example of Sec.~\ref{sec:marginal}.

The physically motivated case sits at the other end of the table. For a
Hagen--Poiseuille tube at fixed volume, $\alpha=1/2$ with $a_e\propto\ell_e^{3/2}$, the state found has $29$ active edges and $\beta=2$, a near-tree backbone on
the same 28-node candidate graph, which is the point of contact between this
formalism and vascular architecture. Its access gain, $15.27$, is more than twice
the largest value reached anywhere else in the table.

\begin{table}[b]
\caption{\label{tab:spatial}%
Length-weighted spatial network, $n=28$ and $m=95$. Access gain is optimized
access divided by equal-share access at the same $\alpha$ and budget; values at
different $\alpha$ are not comparable in physical units. The first row is the
Hagen--Poiseuille fixed-volume law of Sec.~\ref{sec:setup}, $\alpha=1/2$ with
$a_e\propto\ell_e^{3/2}$; the remaining rows use $a_e=\ell_e$. All rows use the
same random start set.}
\begin{ruledtabular}
\begin{tabular}{rlrrrrr}
$\alpha$ & $a_e$ & Active & Cycle & $N_p$ & $N_w$ & Access\\
& & edges & rank & & & gain\\
\hline
0.50 & $\ell_e^{3/2}$ & 29 &  2 & 18.4 & 15.8 & 15.27\\
0.55 & $\ell_e$ & 30 &  3 & 23.2 & 24.5 & 6.07\\
0.75 & $\ell_e$ & 32 &  5 & 22.6 & 23.5 & 2.89\\
0.95 & $\ell_e$ & 60 & 33 & 34.7 & 38.8 & 2.07\\
1.00 & $\ell_e$ & 86 & 59 & 38.4 & 44.8 & 1.99\\
1.25 & $\ell_e$ & 95 & 68 & 36.9 & 56.9 & 1.77\\
2.00 & $\ell_e$ & 95 & 68 & 30.9 & 76.2 & 1.48\\
\end{tabular}
\end{ruledtabular}
\end{table}

\begin{figure*}[t]
\centering
\includegraphics[width=\textwidth]{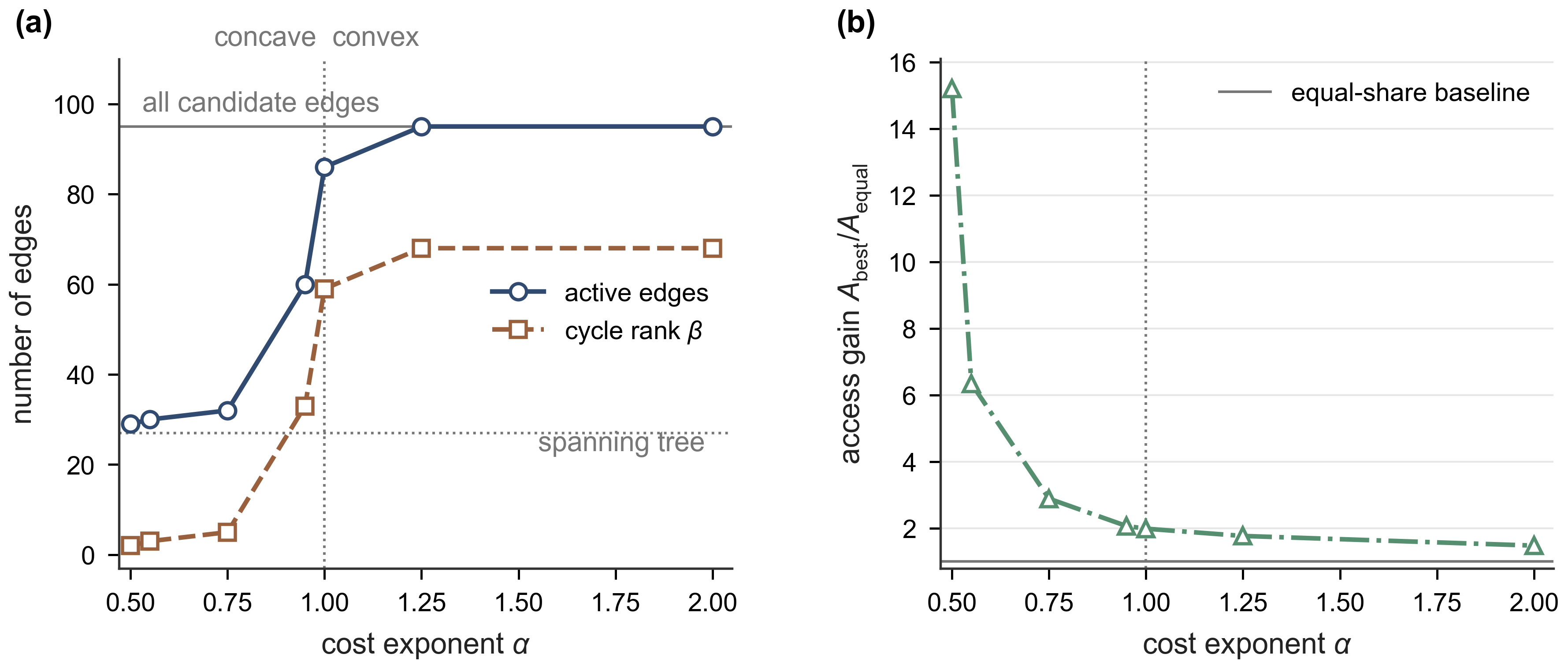}
\caption{\label{fig:spatialgain}%
Support topology and transport improvement in the spatial candidate graph. (a)
Active edges and independent cycles against $\alpha$; to the left of the dotted
line at $\alpha=1$ the cost is concave and a sparse backbone is stabilized, while
convex cost recruits loop redundancy, with the spanning-tree and all-edges
reference lines shown. (b) Access gain
$A_{\rm best}/A_{\rm equal}$, the factor by which optimal allocation beats equal
budget shares, largest in the concave regime where concentrating conductance on a
few channels has the strongest effect.}
\end{figure*}

The two effective supports separate as $\alpha$ grows. They are close in the
concave cases, $23.2$ against $24.5$ at $\alpha=0.55$, and diverge strongly in the
convex ones, $30.9$ against $76.2$ at $\alpha=2$. Convex cost thus spreads the
conductance far more evenly than it spreads the budget, because the
budget-to-conductance map compresses differences when $\alpha>1$. The map
amplifies them when $\alpha<1$, and the ordering does reverse, though only well
below the boundary: at $\alpha=1/2$ the conductance is more concentrated than the
budget that buys it, $N_w=15.8$ against $N_p=18.4$. The access gain
$A_{\rm best}/A_{\rm equal}$ in Fig.~\ref{fig:spatialgain} runs in the opposite
direction, from $15.27$ at $\alpha=1/2$ and $6.07$ at $\alpha=0.55$ to $1.48$ at
$\alpha=2$: allocation is worth most precisely where the state found is sparsest and hardest to find, and therefore most in the physically motivated
fixed-volume case.

\section{The load decides whether loops are worth buying}
\label{sec:loading}
The cost law is not the only control of architecture. What the network is asked to
do matters just as much, and the cleanest way to show this is to change the
loading while holding everything else fixed. On the 20-node geometric graph we
compare the all-pairs objective against a single source--sink pair separated by
six hops.

For the single pair, let $f_e$ be an oriented unit flow from $s$ to $t$. Thomson's
principle writes $r_{st}$ as the minimum of $\sum_e f_e^2/w_e$ over such flows.
Eliminating $W$ at fixed $f$ under Eq.~(\ref{eq:cost}) leaves the problem of
minimizing
\begin{equation}
S(f)=\sum_e a_e^{1/(\alpha+1)}\,|f_e|^{2\alpha/(\alpha+1)},
\label{eq:Sflow}
\end{equation}
the remaining positive factors being independent of $f$. For $0<\alpha<1$ the
exponent $q=2\alpha/(\alpha+1)$ lies below one, so $|f_e|^q$ is concave in $|f_e|$
on the non-negative half-line. The function $S$ is not concave on the full space of
signed flows, and the argument requires the orientation to be fixed first. Any unit
$s$--$t$ flow decomposes into simple path flows plus circulations. Removing the
circulations cannot raise the dissipation functional, because it is increasing in
each $|f_e|$ for $q>0$ and the decomposition lowers $|f_e|$ pointwise; what
remains is an acyclic flow, and adopting the orientation that this acyclic
flow induces on each edge makes all remaining coordinates non-negative. On that
directed polytope $S$ is concave, the polytope is the convex hull of the simple
$s$--$t$ path flows, and a concave function attains its minimum at an extreme
point. Some minimizer is therefore supported on a single path and carries no
cycles.

The structure of this argument is not new. Banavar and co-workers showed that
minimizing a concave function of the link flows at prescribed sources and sinks
selects a spanning tree, by exactly this route: concavity moves the minimum to an
extreme point of the flow polytope~\cite{Banavar2000}, a result then developed for
concave transport costs and loop analysis by Shao and Zhou~\cite{Shao2007}.
Equation~(\ref{eq:Sflow}) is the same mechanism written for a resistance
objective, and the single-pair result here should be read as recovering their
conclusion in our variables rather than as an independent finding.

What differs is the loading. Their optimum is defined at a prescribed flow
pattern, so the concave argument applies directly and looplessness follows for the
whole network. Our objective averages over every node pair, which is not a single
flow problem and admits no such reduction: the concave argument bites only in the
rank-one case $Q=b_{st}b_{st}^{\mathsf T}$, and at finite $n$ the all-pairs states
in Table~\ref{tab:loading} carry up to 32 independent cycles. Two further
differences follow from working with $\Rbar$ rather than with dissipation at fixed
flow. The convex side of the same problem supplies the reference state that makes
Eq.~(\ref{eq:eta}) an absolute score and splits it into $\etatop$ and $\etaw$,
which has no counterpart in the prescribed-flow formulation. And the all-pairs
ensemble leaves room for the size dependence of Sec.~\ref{sec:scaling}, where
whether loops survive is a question about $n$ and not only about the cost
exponent.

On a fixed path $P$, equalizing $\kappat_e$ along it gives
$w_e\propto a_e^{-1/(\alpha+1)}$ and an optimized resistance
\begin{equation}
r_{st}(P)=C_0^{-1/\alpha}\left[\sum_{e\in P}a_e^{1/(\alpha+1)}\right]^{1+1/\alpha},
\label{eq:pathresistance}
\end{equation}
so the optimal $P$ is the shortest path under the transformed edge price
$a_e^{1/(\alpha+1)}$. With equal prices this is the ordinary six-edge shortest
route used in Table~\ref{tab:loading}.

\begin{table}[t]
\caption{\label{tab:loading}%
All-pairs loading against the exact single-pair path on the same geometric
candidate graph, $n=20$ and $m=62$. The path is loopless at every exponent
shown. The all-pairs search uses the same two-stage optimizer and continuation
protocol as the rest of the paper; relative to the earlier pipeline only the $\alpha=0.65$ row moves, to
a lower resistance with one additional active edge.}
\begin{ruledtabular}
\begin{tabular}{rrrrr}
$\alpha$ & Eff. support & Active & Cycles & Edges in exact\\
& (all pairs) & (all pairs) & (all pairs) & single-pair path\\
\hline
0.45 & 19.9 & 21 &  2 & 6\\
0.55 & 21.6 & 23 &  4 & 6\\
0.65 & 24.0 & 27 &  8 & 6\\
0.75 & 26.7 & 33 & 14 & 6\\
0.90 & 32.7 & 43 & 24 & 6\\
0.97 & 33.5 & 51 & 32 & 6\\
\end{tabular}
\end{ruledtabular}
\end{table}

Table~\ref{tab:loading} shows this contrast. Under all-pairs loading the
effective support grows from $19.9$ at $\alpha=0.45$ to $33.5$ at $\alpha=0.97$
and the cycle rank from 2 to 32. The exact single-pair optimum has six active
edges and no cycles at every exponent, and the looplessness holds for all
$0<\alpha<1$, although which path is selected can depend on $\alpha$ when the
coefficients $a_e$ are unequal.

Prescribing a source and a sink removes the mechanism that pays for loops here.
Redundancy is purchased when the network must serve pairs that do not share a
route; for a single pair the concave flow cost selects one path exactly. Read
together with Sec.~\ref{sec:scaling}, the two results point the same way: loops in
the best concave all-pairs states found are contingent, disappearing when the
loading is narrowed to one pair and also, at $\alpha=0.65$, when the system is
simply made larger.

\section{Reading the marginal law on a network one did not design}
\begin{figure*}[t]
\centering
\includegraphics[width=\textwidth]{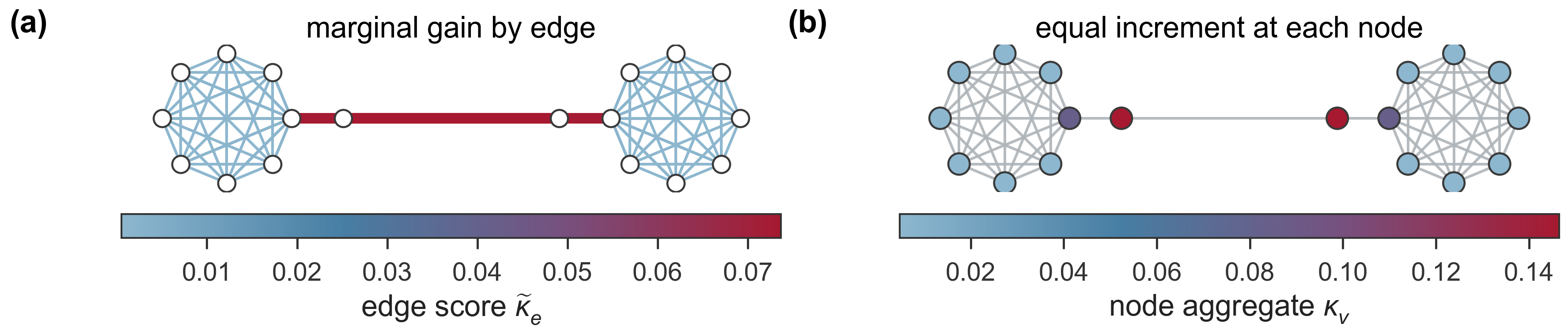}
\caption{\label{fig:barbell}%
Marginal access gain per unit cost in a barbell network with two eight-node
cliques and two intermediate nodes, at $\alpha=2$. Darker and thicker bridge edges
carry the largest sensitivity; the maximum-to-minimum ratio is 144, or $2.16$
orders of magnitude. Panel (a) shows the edge scores. Panel (b) sums the incident
edge gains at each node for the equal-increment direction.}
\end{figure*}

Equation~(\ref{eq:kappa}) can be evaluated on a measured network without
optimizing its topology at all, which is how it would be used in practice.
Figure~\ref{fig:barbell} does this for a barbell of two eight-node cliques joined
by a three-edge bridge path through two intermediate nodes, with unit
conductances, unit edge-price coefficients, and $\alpha=2$. This is a smaller
graph than the 24-node \texttt{barbell\_graph(11,\,2)} of
Table~\ref{tab:convex}. The largest
$\kappat_e$ lie on the bridge, and $\kappat_{\max}/\kappat_{\min}=144$, or $2.16$
orders of magnitude. A uniform network is thus very far from equalizing marginal
value, which is precisely the signal Eq.~(\ref{eq:kappa}) is meant to carry.

Panel (b) reports the aggregate $\kappa_v=\sum_{e\ni v}\kappat_e$. This is a
diagnostic sum of incident edge scores, equivalent to assigning the same
infinitesimal resource increment to every edge at $v$. It is not an intrinsic
derivative of $\eta$ with respect to a node, and it should not be quoted as one.

\section{Discussion}
The exponent $\alpha$ separates two mechanisms that are easy to conflate. For $\alpha>1$ the marginal cost of activating an absent edge vanishes, so every admissible channel receives positive conductance at the optimum and the design problem is entirely one of allocation. For $\alpha<1$ the first increment on a new channel has a divergent marginal cost, so access improves mainly by reinforcing an existing backbone, and topology becomes part of the answer rather than an input. Loops return only when the reduction they buy in the all-pairs effective resistance outweighs the price of opening additional channels.

Section~\ref{sec:scaling} sharpens what ``sparse'' means here. Across the sizes we can reach, the
concave states found use a decreasing fraction of the candidate set, but the number of channels per node changes much less. An active fraction measured at $n=12$ or $20$ will therefore not transfer directly to a larger graph. The behavior of loops is less simple. At $\alpha=0.65$ the cycles present through $n=28$ are gone at all five larger sizes, and seeded continuation at $n=32$ and $40$ does not recover them. Some of the redundancy seen in small concave states may therefore be a finite-size effect rather than a general consequence of the cost law. This does not happen over the accessible range at $\alpha=0.75$, where $\beta/n$ averages $1.64$ over the last four sizes and reaches $1.78$ at $n=64$. If a size-driven crossover exists at that exponent, it lies beyond the present calculation.

The closest earlier comparison is the work of Shao and Zhou~\cite{Shao2007}, who studied concave transport costs and the stability of loops for prescribed sources and sinks. Their objective is tied to that load and to the flows it produces. Here the mean effective resistance averages over every unordered node pair, so one added edge can improve many transport routes at once. The size dependence must therefore be read for this all-pairs ensemble. At
$\alpha=0.65$ the states found carry loops through $n=28$ but become trees from $n=32$ onward. At $\alpha=0.75$, by contrast, $\beta/n$ averages $1.64$ over the last four sizes and is still rising at $n=64$, where it reaches $1.78$. Concave cost alone does not make loops inevitable; their value depends on the loading ensemble and on system size. Corson reached a complementary conclusion for prescribed
loading~\cite{Corson2010}: with a single stationary source configuration the
optimum is a tree, and loops appear once the load fluctuates, so redundancy is
bought by variability in the demand rather than by the cost law. Our all-pairs
objective is a stationary but non-degenerate load, and it sits between his two
cases, which is consistent with loops that are present but contingent on
$\alpha$ and $n$. Kaiser \emph{et al.} showed that the onset of loops in optimal
supply networks is discontinuous, arising through a saddle-node
bifurcation~\cite{Kaiser2020}; the abrupt collapse localized in
Sec.~\ref{sec:scaling} has the same character along the cost exponent. This
complements related studies of optimal transport
architectures~\cite{Bohn2007,Durand2006,Katifori2010,Ronellenfitsch2016}.

The convex index is the part of this that is ready for use on data. Given an observed network and a candidate edge set, $\etaw$ scores the allocation of conductance and $\etatop$ scores the restriction imposed by the wiring, both at the network's own budget. The comparison is only meaningful when weight, cost exponent, geometric coefficient, node set, and transport ensemble carry the same interpretation in the two systems being compared. A correlation matrix, a hydraulic conductance, and an electrical admittance do not belong on one scale merely because each is called a weighted network.

The spatial networks considered here are synthetic. Measured channel geometries will be needed to test the length penalty in a physical system, and the fixed-volume case of Sec.~\ref{sec:spatial} is the natural point of contact with measured vascular architectures~\cite{Ronellenfitsch2019}. More general loading is representable by a symmetric positive-semidefinite demand matrix $Q$ with $Q\mathbf{1}=0$, giving the objective $\operatorname{tr}(L^{+}Q)$; all-pairs loading is the isotropic choice on $\mathbf{1}^{\perp}$ and a single pair is the rank-one case $Q=b_{st}b_{st}^{\mathsf T}$. The perturbation argument of Sec.~\ref{sec:marginal} carries over unchanged, and whether the per-node loop count is controlled by the spectrum of $Q$ is a well-posed question that the two extreme cases here leave open. The concave all-pairs states remain numerical candidates throughout, and every support size, cycle rank, and crossover reported for $\alpha<1$ is a property of the best state found at a stated search budget.

Certifying global optimality on general graphs is a separate problem, and the controls in Sec.~\ref{sec:scaling} bound the risk without removing it. There is, however, a way to recover an absolute statement below $\alpha=1$ without solving the nonconvex problem, and it seems worth stating as a target. Any lower bound $L\le\Rbar^{*}$ obtained independently of the search, whether from a convex relaxation of Eq.~(\ref{eq:optimization}) or from a combinatorial argument on the candidate set, immediately gives $L/\Rbar\le\eta_\alpha\le U/\Rbar$ with $U$ the resistance of any feasible state, in particular the best one found. That is a certified interval rather than a point estimate, it narrows as either bound improves, and it would extend Eq.~(\ref{eq:eta}) to the concave regime where we currently decline to use it. 

A useful relaxation is not obvious, since the difficulty sits in the concave cost rather than in $\Rbar$, which is convex throughout; we have not pursued it here. The optimizer iterations should not be interpreted as a physical adaptation process. A biological or materials model would additionally need a mobility law, kinetic coefficients, and a mechanism linking local observations to conductance changes; adaptive dynamics of that kind, in which conductances respond to the flows they carry, are the subject of a separate line of work~\cite{HuCai2013}.

\section{Conclusion}
A finite resource budget gives a common physical basis for comparing weighted transport networks. Where the cost is convex, the mean effective resistance has a unique optimal allocation, which turns a normalized efficiency into a genuine measurement and factors it into an allocation part and a topological part. Its gradient is a squared biharmonic distance, giving the marginal worth of every channel and a marginal-value balance of Murray type in which the balanced quantity is a property of the whole network.
Concave cost changes both the allocation and the active topology. Sparse supports and competing local minima appear, while the uniform star and complete graph provide only two analytic reference states. For $\alpha=0.75$ and $0.90$, the support found numerically grows approximately linearly with $n$ over the sizes examined. The behavior of loops is less uniform. Resolving the loop-collapse exponent $\alpha^{*}(n)$ on a grid of step $0.005$ shows it rising at every size from $0.555$ at $n=8$ to $0.675$ at $n=64$, between $0.15$
and $0.16$ below $\alpha_c(n)$ throughout, with a single and abrupt transition at every size. The change of character reported at the fixed exponent $\alpha=0.65$ between $n=28$ and $n=32$ is this rising boundary being crossed, and at that point no looped state is recovered by seeded continuation in $n$, by continuation downward in $\alpha$, or by opening any closed channel with a finite jump. Loading provides a separate control: for a single prescribed pair, Thomson's principle gives a loopless optimum throughout $0<\alpha<1$. Cost convexity and loading geometry together set both the access a network can reach and the architecture that delivers it.

\begin{acknowledgments}
This work was supported by DGAPA-PAPIIT under Grant No. IN234029 and by SECIHTI under Grant No. CBF-2025-G-789.
\end{acknowledgments}

\section*{Data availability}
All results reported here are generated by the executable notebook provided as
Supplemental Material, which also documents the software environment used.
Supplemental Table~S1 lists the 44 states of the size scan.

Claude Opus 5 (Anthropic) was used for language editing, for assistance with
code scripting, and to rerun part of the calculations in a second software
environment, as recorded in the notebook. All results were verified by the
authors, who take full responsibility for the content.

\bibliography{references}

@article{Klein1993,
  author  = {Klein, Douglas J. and Randi{\'c}, Milan},
  title   = {Resistance distance},
  journal = {J. Math. Chem.},
  volume  = {12},
  number  = {1},
  pages   = {81--95},
  year    = {1993},
  doi     = {10.1007/BF01164627}
}

@article{Latora2001,
  author  = {Latora, Vito and Marchiori, Massimo},
  title   = {Efficient Behavior of Small-World Networks},
  journal = {Phys. Rev. Lett.},
  volume  = {87},
  pages   = {198701},
  year    = {2001},
  doi     = {10.1103/PhysRevLett.87.198701}
}

@book{Newman2010,
  author    = {Newman, M. E. J.},
  title     = {Networks: An Introduction},
  publisher = {Oxford University Press},
  address   = {Oxford},
  year      = {2010}
}

@article{Ghosh2008,
  author  = {Ghosh, Arpita and Boyd, Stephen and Saberi, Amin},
  title   = {Minimizing Effective Resistance of a Graph},
  journal = {SIAM Rev.},
  volume  = {50},
  number  = {1},
  pages   = {37--66},
  year    = {2008},
  doi     = {10.1137/050645452}
}

@article{Banavar2000,
  author  = {Banavar, Jayanth R. and Colaiori, Francesca and
             Flammini, Alessandro and Maritan, Amos and Rinaldo, Andrea},
  title   = {Topology of the Fittest Transportation Network},
  journal = {Phys. Rev. Lett.},
  volume  = {84},
  number  = {20},
  pages   = {4745--4748},
  year    = {2000},
  doi     = {10.1103/PhysRevLett.84.4745}
}

@article{Bohn2007,
  author  = {Bohn, Steffen and Magnasco, Marcelo O.},
  title   = {Structure, Scaling, and Phase Transition in the Optimal Transport Network},
  journal = {Phys. Rev. Lett.},
  volume  = {98},
  pages   = {088702},
  year    = {2007},
  doi     = {10.1103/PhysRevLett.98.088702}
}

@article{Shao2007,
  author  = {Shao, Zhen and Zhou, Haijun},
  title   = {Optimal Transportation Network with Concave Cost Functions: Loop Analysis and Algorithms},
  journal = {Phys. Rev. E},
  volume  = {75},
  pages   = {066112},
  year    = {2007},
  doi     = {10.1103/PhysRevE.75.066112}
}

@article{Durand2006,
  author  = {Durand, Marc},
  title   = {Architecture of Optimal Transport Networks},
  journal = {Phys. Rev. E},
  volume  = {73},
  pages   = {016116},
  year    = {2006},
  doi     = {10.1103/PhysRevE.73.016116}
}

@article{Durand2007,
  author  = {Durand, Marc},
  title   = {Structure of Optimal Transport Networks Subject to a Global Constraint},
  journal = {Phys. Rev. Lett.},
  volume  = {98},
  pages   = {088701},
  year    = {2007},
  doi     = {10.1103/PhysRevLett.98.088701}
}

@article{Katifori2010,
  author  = {Katifori, Eleni and Sz{\"o}ll{\H{o}}si, Gergely J. and
             Magnasco, Marcelo O.},
  title   = {Damage and Fluctuations Induce Loops in Optimal Transport Networks},
  journal = {Phys. Rev. Lett.},
  volume  = {104},
  number  = {4},
  pages   = {048704},
  year    = {2010},
  doi     = {10.1103/PhysRevLett.104.048704}
}

@article{Ronellenfitsch2016,
  author  = {Ronellenfitsch, Henrik and Katifori, Eleni},
  title   = {Global Optimization, Local Adaptation, and the Role of Growth in Distribution Networks},
  journal = {Phys. Rev. Lett.},
  volume  = {117},
  pages   = {138301},
  year    = {2016},
  doi     = {10.1103/PhysRevLett.117.138301}
}

@article{Murray1926,
  author  = {Murray, Cecil D.},
  title   = {The Physiological Principle of Minimum Work. {I}. {T}he Vascular System and the Cost of Blood Volume},
  journal = {Proc. Natl. Acad. Sci. U.S.A.},
  volume  = {12},
  number  = {3},
  pages   = {207--214},
  year    = {1926},
  doi     = {10.1073/pnas.12.3.207}
}

@article{Lonardi2021,
  author  = {Lonardi, Alessandro and Facca, Enrico and Putti, Mario and
             De Bacco, Caterina},
  title   = {Designing Optimal Networks for Multicommodity Transport Problem},
  journal = {Phys. Rev. Research},
  volume  = {3},
  pages   = {043010},
  year    = {2021},
  doi     = {10.1103/PhysRevResearch.3.043010}
}

@article{Lipman2010,
  author  = {Lipman, Yaron and Rustamov, Raif M. and Funkhouser, Thomas A.},
  title   = {Biharmonic Distance},
  journal = {ACM Trans. Graph.},
  volume  = {29},
  number  = {3},
  pages   = {27},
  year    = {2010},
  doi     = {10.1145/1805964.1805971}
}

@inproceedings{Black2024,
  author    = {Black, Mitchell and Lin, Lucy and Wong, Weng-Keen and
               Nayyeri, Amir},
  title     = {Biharmonic Distance of Graphs and its Higher-Order Variants: Theoretical Properties with Applications to Centrality and Clustering},
  booktitle = {Proceedings of the 41st International Conference on Machine Learning},
  series    = {Proceedings of Machine Learning Research},
  volume    = {235},
  pages     = {4077--4102},
  year      = {2024},
  publisher = {PMLR},
  url       = {https://proceedings.mlr.press/v235/black24a.html}
}

@article{Bejan2004,
  author  = {Bejan, Adrian and Lorente, Sylvie},
  title   = {The Constructal Law and the Thermodynamics of Flow Systems with Configuration},
  journal = {Int. J. Heat Mass Transfer},
  volume  = {47},
  number  = {14--16},
  pages   = {3203--3214},
  year    = {2004},
  doi     = {10.1016/j.ijheatmasstransfer.2004.02.007}
}

@article{Bejan2010,
  author  = {Bejan, Adrian and Lorente, Sylvie},
  title   = {The Constructal Law of Design and Evolution in Nature},
  journal = {Philos. Trans. R. Soc. B},
  volume  = {365},
  number  = {1545},
  pages   = {1335--1347},
  year    = {2010},
  doi     = {10.1098/rstb.2009.0302}
}

@article{Corson2010,
  author  = {Corson, Francis},
  title   = {Fluctuations and Redundancy in Optimal Transport Networks},
  journal = {Phys. Rev. Lett.},
  volume  = {104},
  pages   = {048703},
  year    = {2010},
  doi     = {10.1103/PhysRevLett.104.048703}
}

@article{Kaiser2020,
  author  = {Kaiser, Franz and Ronellenfitsch, Henrik and Witthaut, Dirk},
  title   = {Discontinuous transition to loop formation in optimal supply networks},
  journal = {Nat. Commun.},
  volume  = {11},
  pages   = {5796},
  year    = {2020},
  doi     = {10.1038/s41467-020-19567-2}
}

@article{HuCai2013,
  author  = {Hu, Dan and Cai, David},
  title   = {Adaptation and Optimization of Biological Transport Networks},
  journal = {Phys. Rev. Lett.},
  volume  = {111},
  pages   = {138701},
  year    = {2013},
  doi     = {10.1103/PhysRevLett.111.138701}
}

@article{Ronellenfitsch2019,
  author  = {Ronellenfitsch, Henrik and Katifori, Eleni},
  title   = {Phenotypes of Vascular Flow Networks},
  journal = {Phys. Rev. Lett.},
  volume  = {123},
  pages   = {248101},
  year    = {2019},
  doi     = {10.1103/PhysRevLett.123.248101}
}

\end{document}